\documentclass[preprint,preprintnumbers,prd,amsmath,amssymb,nofootinbib,tightenlines]{revtex4}
\usepackage{graphicx}
\usepackage{dcolumn}
\usepackage{bm}

\begin{document}
\preprint{SLAC-PUB-14812} 

\title{\boldmath%
Pair Production Constraints on Superluminal Neutrinos Revisited 
\unboldmath}

\author{Stanley J. Brodsky}

\affiliation{
SLAC National Accelerator Laboratory, Stanford University, Stanford,
California 94309
}

\author{Susan Gardner}

\affiliation{
Department of Physics and Astronomy, University of Kentucky, 
Lexington, KY 40506-0055
}

\date{\today}

\begin{abstract}
We revisit the pair creation constraint on superluminal neutrinos considered by 
Cohen and Glashow in 
order to clarify which types of superluminal models are constrained. We show 
that a model in which the superluminal neutrino is effectively light-like 
can evade the Cohen-Glashow constraint. 
\end{abstract}

\maketitle

Cohen and Glashow (CG) have shown that the vacuum lepton  pair  production process 
($\nu \to \nu e^+ e^-$) acts as an efficient energy loss mechanism for 
propagating superluminal neutrinos~\cite{CGprl}; the observation
of neutrinos in the OPERA experiment thus appears to exclude
the superluminal interpretation of their velocities~\cite{opera}. 
However, as we shall show, the requirement of energy-momentum conservation 
does not limit the possibility that the superluminal neutrino is effectively light-like.

The group velocity $v_{\rm G}$ of a particle with energy $E$ and momentum 
$\mathbf{p}$ is given by 
$\mathbf{v}_{\rm G}= dE/d\mathbf{p}$.
Assuming a dispersion relation of form 
$E= \sqrt{ |\mathbf{p}|^2 c'^2 + m'^2 c'^4}$~\cite{Coleman:1998ti,Coleman:1997xq}, where $m'c'^2 = m c^2$
and $m$ is the particle mass, 
yields $\mathbf{v}_{\rm G}/c = (|\mathbf{p}|c/E) (c'/ c)^2$. 
Thus a particle can become superluminal, namely,  it can 
attain $v_{\rm G} > c$, where $c$ 
is the measured speed of light in a terrestrial vacuum~\cite{hall}, in two different ways:
\begin{enumerate}
\item
if $c' /c > 1$ with $|\mathbf{p}| c /E \lesssim 1$ {\em or}
\item 
if $|\mathbf{p}|c >  E$ and $c'=c$.
\end{enumerate}
Forming $p=(E,c\mathbf{p})$, so that 
$p^2 = E^2 - c^2|\mathbf{p}|^2$, then if $v_{\rm G} > c$ the superluminal particle 
is timelike in the former case and spacelike in the latter case. 
Searches for particles obeying the latter possibility --- ``tachyons''~\cite{feinberg} ---  have been
conducted previously without success~\cite{Baltay:1970se}. 

The observation of neutrino oscillations sharply constrains 
superluminal neutrinos~\cite{CGprl}, 
and severe limits exist on the velocity difference of neutrinos of different 
species~\cite{Coleman:1998ti,Coleman:1997xq,CGprl,fargion}. 
Thus superluminality, 
if it exists, must pertain to all three neutrino mass eigenstates. 
CG assume a superluminal neutrino by choosing $c' > c$; thus all neutrino mass eigenstates are placed 
on the same footing. 

If one assumes that the neutrino mass matrix is Hermitian~\cite{Coleman:1998ti,Coleman:1997xq}, then 
the possibility of a spacelike superluminal neutrino~\cite{feinberg} is excluded 
from the onset. If one relaxes the Hermiticity assumption, one  
runs afoul of the observation of neutrinos propagating from distant sources. 
Moreover, 
a spacelike superluminal neutrino necessarily becomes {\em more} superluminal as its energy decreases,
although OPERA observes no significant energy dependence in comparing the speed of neutrinos 
with an average energy of 13.9 and 42.9 GeV~\cite{opera}. 
Thus one can safely eliminate spacelike superluminal neutrinos as a possibility~\cite{sterile}. 

The mass scale of neutrinos is sufficiently well-known~\cite{mainz,arXiv:1006.5276} to establish 
that any effect from nonzero neutrino masses 
at OPERA energies is negligible~\cite{opera}; in fact,  since
$E\gg m' c'^2$ with $c' >c$,  then $E \simeq c'|\mathbf{p}|$. Since CG define
$p=(E, c\mathbf {p})$, then $p^2 = (c'^2 - c^2)|\mathbf{p}|^2 >  0$ for finite $|\mathbf{p}|$, and their 
superluminal neutrinos are timelike. Alternatively, one can define $p=(E,c'\mathbf{p})$, so that 
the superluminal neutrino 
is effectively lightlike at high energies, $p^2 = m'^2 c'^4 \simeq 0$ ~\cite{SG}. 
Note that $p^2$ in this latter case remains a Lorentz scalar. Indeed $c$ does not appear explicitly in $p$, 
and Lorentz symmetry is not manifestly broken 
--- $c'$ can be taken to be the causal velocity. 
Nevertheless, in this scenario, the neutrino 
can {\it appear} superluminal compared to the propagation of photons; this is because 
light can suffer interactions in matter which the neutrino does not --- this is already true
in the Standard Model (SM). 

The vacuum Cherenkov-like process which CG find of greatest importance is 
$\nu (p_1) \to \nu (p_2)  + e^+ (l_e') + e^- (l_e)$. Energy-momentum
conservation implies $p_1^2 = p_2^2 + 2 p_2 \cdot (l_e + l_e') + (l_e + l_e')^2$, 
where $(l_e + l_e')^2 >0$ and $p_2 \cdot (l_e + l_e')\gtrsim 0$. Thus
this process cannot occur for effectively lightlike models in which the four momentum squared 
does not change $p_1^2 = p_2^2$, such as in Ref.~\cite{SG}. 
In contrast, a superluminal neutrino as considered by CG has a value of $p^2$ which depends on energy, 
so that $p_1^2 - p_2^2 > 0 $ can occur, 
and the process is indeed allowed. For such neutrinos vacuum Cherenkov processes
act as a new source of neutrino decay.

Pair production will of course occur in matter even if vacuum pair production is forbidden 
--- the transfer of 3-momentum to a target or medium is needed to realize 
energy-momentum conservation, as has been previously studied in the context of pair
production in the Coulomb field of a nucleus of charge $Ze$, 
e.g., $\mu^+ + Z  \to \mu^+ + e^- + e^+ + Z $ in 
quantum electrodynamics~\cite{Brodsky:1966vh,Bjorken:1966kh}. In the model of
Ref.~\cite{SG} pair production associated with a propagating 
superluminal neutrino mimics that of SM neutrinos in ordinary matter --- it 
shares all the same diagrams, except that $c$ is replaced by $c'$ throughout. 
Under this global replacement, $\pi$ decay and cosmic ray constraints also do not operate~\cite{gonzalez,bi}:
Lorentz symmetry is not broken at the level of particle interactions. 
Pair bremsstrahlung is much less effective in matter; Standard Model neutrinos do shed
such radiation, but Earth-crossing neutrinos are both expected and observed~\cite{earth}. 

In the model of Ref.~\cite{SG} $c' > c$ is realized through 
photon interactions with a dark sector which make $c$ smaller than the causal velocity $c'$, 
so that additional diagrams do appear in that case vis-a-vis final-state interactions
of the charged leptons with the dark sector; these are, however, small modifications as 
such couplings must be much weaker than the usual electromagnetic coupling $e$. 
These calculations do not reference the value of $c$ per se. 
The propagation speed of light --- or of electrons or of other particles with
suitable generalization --- in a medium, is fixed by the real part 
of the forward Compton amplitude in that medium: a different calculation altogether. 
Thus the possibility of lightlike superluminal neutrinos is not excluded by the observation
of neutrinos at OPERA energies and beyond. 

An additional constraint on neutrino superluminality comes from the observation of 
neutrinos from SN 1987A: the close  detection in time of the neutrino and optical 
bursts yields a neutrino-light velocity comparison test, namely, of 
$v_\nu - v_\gamma < 2 \times 10^{-9}$~\cite{longo}. 
This constraint does assume that the neutrinos and photons were emitted at the same time. Since 
supernova neutrinos are of ${\cal O}(10 \,\hbox{MeV})$ in energy, 
 reconciling this result with OPERA evidently requires an energy-dependent
neutrino dispersion relation~\cite{arXiv:1109.4980,arXiv:1109.5172,fargion,arXiv:1109.5682,arXiv:1110.1853}. 
Indeed, the generality of the CG constraint in eliminating superluminal 
neutrino models 
rests on the inclusion of the SN 1987A constraint as well~\cite{villante,evslin}. 
Note, however, that the  assumption of contemporary emission of the neutrino and optical bursts 
is not firmly established; it relies on 
an understanding of the SN mechanism --- which we do not 
possess. 
Realistic studies of neutrino transport reveal that 
considerations such as convection, rotation, and magnetic fields 
first investigated in the 1990's are in some measure necessary to 
realize the explosion of a star~\cite{astro-ph/0005366}, and the implementation of this  program 
is ongoing~\cite{arXiv:0907.4043}. 
Moreover, observations of SN 1987A show it to be an outlier 
in many respects~\cite{pod}. The neutrinos need not come from the initial core collapse; 
they can be emitted at a later stage from the edge of an accretion disk around a forming 
black-hole~\cite{wmac,nagataki,mclsur}. 
If the neutrinos were not emitted contemporaneously with the light, there is a coincidence in their
close temporal detection ---thus in order to explain OPERA, one  requires fine-tuning at 
order ${\cal O}(8\times 10^{-5})$,
which is certainly very improbable but not impossible. 

In summary, any model for which the CG pair production process operates is excluded because 
such timelike neutrinos would not  be detected by OPERA or other experiments. 
However, a superluminal neutrino which is effectively lightlike with fixed $p^2$~\cite{SG} 
can evade the Cohen-Glashow constraint because of energy-momentum conservation.
The coincidence involved in explaining the SN1987A constraint certainly makes such a picture 
improbable --- but it is still intrinsically {\em possible}.  
The lightlike model is appealing
in that it does not violate Lorentz symmetry in particle interactions, 
although one would expect Hughes-Drever tests to turn up a violation eventually~\cite{SG}.
Other evasions of the CG constraints are also possible; perhaps, e.g., 
the neutrino takes a ``short cut'' through extra dimensions~\cite{weiler,pak}, or 
suffers anomalous acceleration in matter~\cite{monopole}, or that Lorentz symmetry
is ``deformed''~\cite{arXiv:1111.4994,arXiv:1111.6330}. 

Irrespective of the OPERA result, 
Lorentz-violating interactions~\cite{NBI-HE-82-30,Colladay:1998fq} remain possible, 
and ongoing experimental investigation of such possibilities should continue. 

This work is supported, in part, by the U.S. Department of Energy under
contracts DE--FG02--96ER40989 and DE--AC02--76SF00515.

\end{document}